\documentclass[reprint,amsmath,amssymb,aps,prb,superscriptaddress]{revtex4-2}
\usepackage{graphicx}% Include figure files
\usepackage{dcolumn}% Align table columns on decimal point
\usepackage{bm}% bold math
\usepackage{color}
\usepackage{blindtext}
\usepackage{float}
\usepackage{hyperref}
\usepackage{amsmath}
\usepackage{enumerate}
\usepackage{changes}
\hypersetup{colorlinks=true, allcolors=blue}
\begin{document}
\preprint{APS/PRB}
\title{
Classical-to-Quantum Crossover in 2D TMD Field-Effect Transistors: A First-Principles Study via Sub-10 nm Channel Scaling Beyond Boltzmann Tyranny
}% 
\author{Yu-Chang Chen}
\email{Contact author: yuchangchen@nycu.edu.tw}
%\homepage{http://www.nycu.edu.tw/~chenyc}
\affiliation{ 
Department of Electrophysics, National Yang Ming Chiao Tung University, 1001 Daxue Rd., Hsinchu City 300093, Taiwan%\\This line break forced with \textbackslash\textbackslash
}%
\affiliation{Center for Theoretical and Computational Physics, National Yang Ming Chiao Tung University.}%Lines break automatically or can be forced with \\
\author{Chia-Yang Ling}%
\affiliation{ 
Department of Electrophysics, National Yang Ming Chiao Tung University, 1001 Daxue Rd., Hsinchu City 300093, Taiwan%\\This line break forced with \textbackslash\textbackslash
}%
\author{Ken-Ming Lin}%
\affiliation{ 
Department of Electrophysics, National Yang Ming Chiao Tung University, 1001 Daxue Rd., Hsinchu City 300093, Taiwan%\\This line break forced with \textbackslash\textbackslash
}%
%\affiliation{
% First affiliation for this author
%}%
%\affiliation{
% second institution for this author
%\affiliation{%

\date{\today}% It is always \today, today,

\begin{abstract}
Scaling field-effect transistors (FETs) into the sub-10-nm regime fundamentally alters the transport mechanism, challenging long-standing design rules. This study investigates monolayer Pt–WSe$_2$–Pt FETs with channel lengths from 12 nm to 3 nm, quantifying the competition between semiclassical thermionic current and quantum tunneling. We show that quantum transport, as described by the Landauer formula, asymptotically approaches classical thermionic emission in the long-channel and high-temperature limit, in accordance with Richardson's law. A competition parameter $\zeta$ cleanly delineates the semiclassical-to-quantum transition, and two characteristic temperatures emerge: $T_{\text{op}}$ (minimizing $J_{\text{OFF}}$) and $T_{\text{c}}$ (thermionic onset). For $L_{\text{ch}}<9$ nm, $T_{\text{op}}<300$ K and $J_{\text{OFF}}$ is tunneling-dominated; the 3 nm device remains tunneling-dominated up to 500 K and achieves a subthreshold swing overcoming Boltzmann tyranny via the steep slope of $\tau(E)$. However, the short-channel effect also generates leakage current and makes the transistor difficult to turn off. For $L_{\text{ch}} \geq 9$ nm, $T_{\text{op}}>300$ K and $J_{\text{OFF}}$ is thermionic-dominated, and the subthreshold swing approaches ($\text{Boltzmann tyranny}/\alpha_{\text{in}}$). Consequently, the ideal channel length for 2D FETs is $L_{\text{ch}} \approx 10$ nm. These results provide criteria for selecting the optimal operating temperature and gate-voltage windows in miniaturizing 2D FETs, and pinpoint the crossover at which quantum tunneling current becomes comparable to semiclassical thermionic emission.
\end{abstract}

%Use showkeys class option if keyword
\maketitle

%\tableofcontents

\section{\label{sec:intro}Introduction}

Shrinking transistor size has been a constant challenge for the semiconductor industry and is a key driver behind the advancement of computing technology~\cite{Radosavljevic_IEEE_2022}. Smaller transistors enable more transistors to be packed onto a single chip, thereby increasing computing power within the same physical footprint~\cite{Keyes_IEEE_2006}. They also offer faster switching speeds, which enhance the performance of electronic devices; lower power consumption, which extends battery life; reduced size, which is ideal for portable devices; and lower manufacturing costs, making electronics more affordable.
 
The performance of field-effect transistors (FETs) is typically characterized by key metrics such as the OFF current, ON/OFF current ratio, and subthreshold swing (SS). In the silicon-based semiconductor industry, scaling down the channel length of transistors from the micrometer scale to sub-10 nm introduces significant challenges related to material properties and changes in transport mechanisms. Specifically: (1) the carrier mobility of silicon as a channel material decreases sharply; and (2) strong quantum tunneling effects increase leakage currents, thereby degrading OFF-state performance \cite{Ieong_Science_2004,Tze-Chiang_IEEE_EXPLORE_2009}.

Despite these challenges, the semiconductor industry continues to push the limits of transistor miniaturization \cite{,Natelson_PhysicsWorld_2009}. Ongoing research and development efforts are focused on overcoming these limitations and exploring new technologies to enable further scaling. Among various alternatives, two-dimensional layered TMDs have emerged as a particularly promising class of semiconductor as channel materials due to their unique electronic and structural properties \cite{Duan_Chem_Rev_2024,mezeli_nat_rev_matt_2017}. 
The carrier mobility in TMDs has been improved to exceed that of silicon \cite{Akinwande_Nature_2019}, and further enhancement may be achieved through lattice distortion \cite{Ng_Nat_Elct_2022}.

Two-dimensional (2D) materials can be integrated into three-dimensional (3D) architectures to enhance transistor density on a chip \cite{Lee_NanoLett_2024}. However, developing robust and reliable manufacturing processes for 2D materials and devices presents a significant challenge. Most 2D transition TMD-based junctions are fabricated using a top-contact geometry, with channel lengths that exceed the electron–phonon mean free path \cite{Radisavljevic_Nat_nanoten_2011,Najmaei_ACSNano_2014,Liu_NanoLett_2016}, resulting in higher resistance and increased heat dissipation due to Joule heating.

In addition, van der Waals interactions at the interface between TMD materials and metal electrodes often result in substantial contact resistance due to the formation of a Schottky barrier~\cite{Novoselov_Science_2016,Somvanshi_PRB_2017,Wu_Nat_Rev_Mat_2023}. This elevated contact resistance reduces the ON current and weakens the overall transistor signal. Consequently, optimizing the metal–TMD interface to minimize contact resistance is critical for improving device performance\cite{Boehm_NanoLett_2023}. Significant efforts, such as doping~\cite{Khalil_ACS_Appl_Mat_Int_2015,Jiang_nat_elect_2024} and semiconducting contact~\cite{Li_Nat_2023}, have been devoted to addressing this issue; however, contact resistance remains a major bottleneck in 2D transistor technology.

Due to manufacturing complexity, the fabrication of 2D TMD junctions with edge-contact geometry is even more challenging. However, edge contacts offer potential advantages, as chemical bonding between the TMD channel material and the metal electrode may reduce the Schottky barrier and lower the contact resistance.

Moreover, as the channel length is scaled down to sub-10 nm dimensions, the electron transport mechanism can undergo a transition from classical to quantum behavior. In the quantum transport regime, the classical limit of the subthreshold swing (SS), referred to as the Boltzmann tyranny (BT), can potentially be surpassed, opening new opportunities for field-effect transistors (FETs) to operate efficiently at low temperatures in the quantum regime~\cite{Nadeem_NanoLett_2021}. To the best of our knowledge, the effects of the classical-to-quantum crossover on the performance of 2D TMD FETs, particularly at sub-10 nm scales, have not yet been systematically investigated. 

\begin{figure*}
\centering
\includegraphics[width=0.7\textwidth]{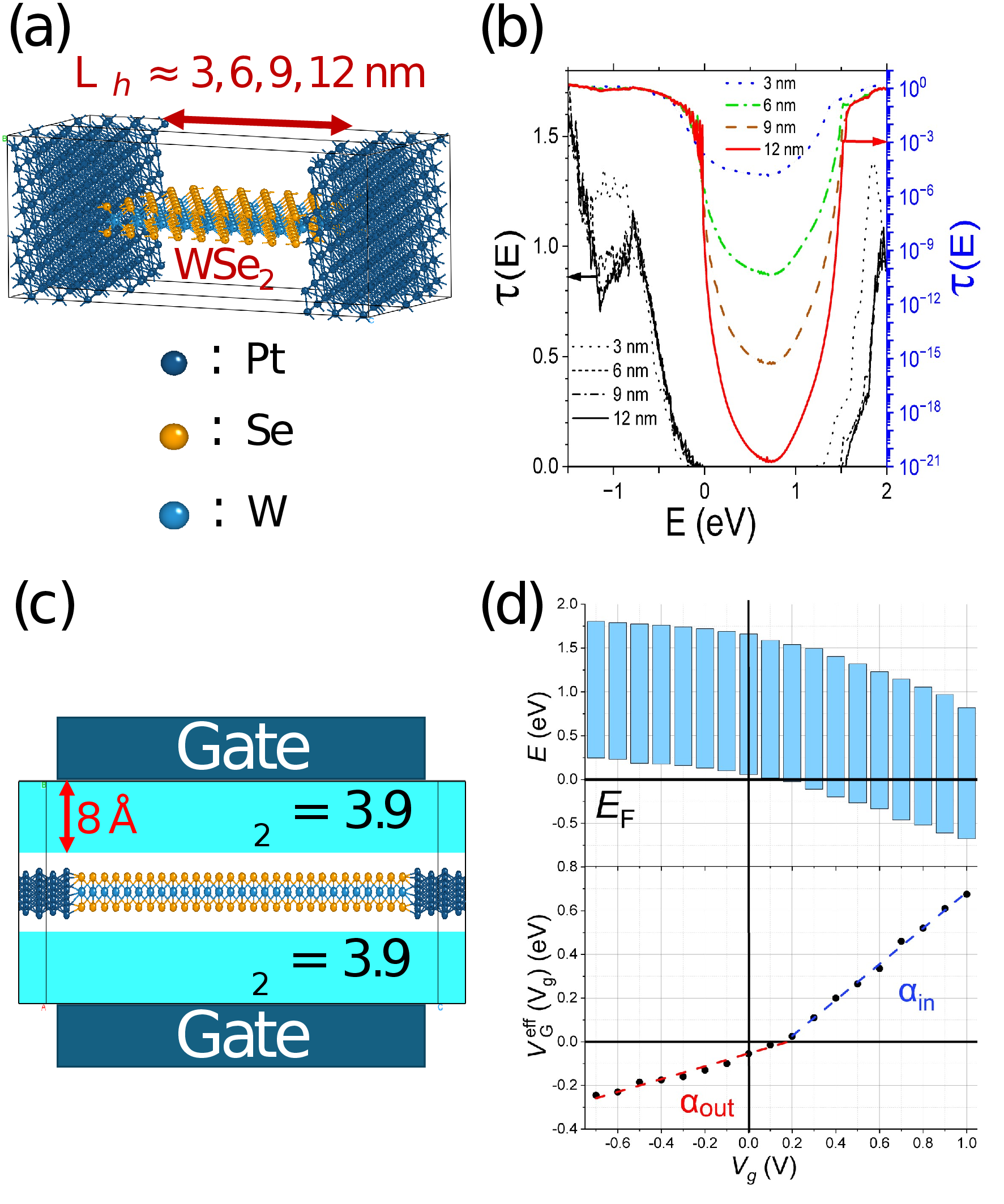} 
\caption{
(a) Schematic of the Pt–WSe$_2$–Pt nanojunction. (b) Transmission coefficients $\tau(E)$ computed from NEGF-DFT (NanoDCAL) for $L_{\mathrm{ch}} = 3, 6, 9,$ and $12$ nm on linear (left axis) and $\log_{10}$ (right axis) scales. (c) Schematic of the gate architecture of the nanojunction as a field-effect transistor with an EOT of 8 \AA (dielectric constant 3.9). (d) The upper panel shows the transmission band gap $(E_V , E_C)$ (vertical blue bars) shifted by the gate voltage $V_g$, with the reference energy being the chemical potential $\mu = 0$ at $V_g=0$ (the black horizontal line). The lower panel illustrates the correlation between $V_g$ and $V_{\mathrm{G}}^{\mathrm{eff}}(V_g)$. Refer to Ref.[\citenum{FET-AlN}].
}
\label{Fig:scheme}
\end{figure*}

%ADD in respond to Comment 1.3 and 2.4:
Bohr’s correspondence principle states that, in the limit of large quantum numbers, the predictions of quantum mechanics converge to those of classical physics. In this spirit, a highly excited system behaves increasingly according to classical mechanical laws. Building on the transmission formalism $\tau(E)$ developed in Ref.~[\citenum{EMT-PW}], we demonstrate an explicit manifestation of the correspondence principle for electrical current in the ballistic transport regime, where the channel length is shorter than the electron–phonon mean free path. In this study, we show that the quantum-mechanical current described by the Landauer formula asymptotically approaches the classical thermionic-emission current in the semiclassical limit. This quantum–classical correspondence motivates the definition of a quantum–classical competition parameter, $\zeta$, which we introduce to quantify and analyze the transition between quantum tunneling and thermionic emission. The connection between the Landauer formula and the classical Richardson law emerges naturally from a rigorous theoretical derivation, from which the parameter $\zeta$ is also systematically established.

The interplay between classical and quantum transport—and its influence on scaling two-dimensional (2D) field-effect transistors (FETs) to the sub-10-nm regime—remains insufficiently understood. When the channel length $L_{\mathrm{ch}}$ is shorter than the electron–phonon mean free path~\cite{Guo_NJPhys_2019,Bai_NanoSelect_2022,Lee_NatCom_2021}, electron–phonon scattering can be neglected and transport is ballistic. This study investigates monolayer Pt–WSe$_2$–Pt nanojunctions designed as FETs with channel lengths of 3, 6, 9, and 12 nm, as illustrated in Fig.~\ref{Fig:scheme}(a). Transmission coefficients $\tau(E)$ are calculated utilizing NEGF-DFT (NanoDCAL), as shown in Fig.~\ref{Fig:scheme}(b). The gate-dependent currents are assessed using the Landauer formalism in conjunction with an effective gate model, $V_{\mathrm{G}}^{\mathrm{eff}}(V_g)$, described in Sec.~IIC. The gate architecture is configured with an equivalent oxide thickness (EOT) of 8 \AA, represented as a dielectric layer having a relative permittivity of $\varepsilon_r = 3.9$, as depicted in Fig.~\ref{Fig:scheme}(c). $\tau(E)$ is shifted by the applied gate voltage $V_g$ such that the transmission band gap $(E_V, E_C)$ relative to the chemical potential $\mu$ at $V_g=0$, as shown in the upper panel of Fig.~\ref{Fig:scheme}(d). The correlation between $V_g$ and $V_{\mathrm{G}}^{\mathrm{eff}}(V_g)$ is depicted in the lower panel of Fig.~\ref{Fig:scheme}(d) \cite{FET-AlN}.

Although the NEGF-DFT framework, combined with the Landauer formalism, is fundamentally a quantum transport approach, we demonstrate in Subsec.~\ref{sec:asymptotic} that the resulting quantum transport current asymptotically reproduces the classical thermionic emission current in the appropriate limit. The correspondence principle enables a rigorous investigation of the crossover in transport mechanisms—from quantum tunneling to classical thermionic emission—as the channel length increases and the temperature rises in scaled field-effect transistors (FETs). This combined framework also enables a comprehensive investigation of FET performance, including the OFF current, ON/OFF current ratio, subthreshold swing, and the operational range of efficient performance across gate voltages $V_g$ from –1.5 to 1.5 V and temperatures ranging from 100 to 500 K, as the channel length is scaled from 12 nm down to 3 nm. Notably, we demonstrate that the shortest nanojunctions (3 nm) exhibit pronounced quantum transport characteristics, with a subthreshold swing (SS) that surpasses the Boltzmann tyranny (BT).

From the analysis of the OFF current and ON/OFF ratio, two characteristic temperatures, $T_{\text{op}}$ and $T_{\text{c}}$ are observed: (1) $T_{\text{op}}$ denotes at which the OFF current reaches its minimum, indicating the most efficient suppression of leakage current in the OFF state; (2) $T_{\text{c}}$ denotes the temperature above which thermionic emission overwhelmingly dominates, and the subthreshold swing approaches  $\mathrm{BT}/\alpha_{\mathrm{in}}$.

\section{Theory}\label{sec:theory}
\subsection{VASP (DFT)}

The geometries of the platinum electrodes, the WSe$_2$ monolayer, and the combined Pt–WSe$_2$–Pt nanojunctions, depicted in Fig.~\ref{Fig:scheme}(a), were optimized using the Vienna Ab-initio Simulation Package (VASP). VASP employed the accurate full-potential projected augmented wave method with a plane-wave basis to resolve the Kohn-Sham problem. \cite{VASP1, VASP2, VASP3, VASP4} In these Density Functional Theory (DFT) computations, the Perdew-Burke-Ernzerhof functional (PBE), a variant of Generalized Gradient Approximation (GGA), is employed to account for many-body effects within the effective single-particle picture. All calculations utilized a grid size of $0.016$~\AA$^{-1}$ in reciprocal space and a plane-wave energy cutoff of $400$~eV. The criterion for terminating the electronic self-consistent iterations was set at $1.0\times 10^{-4}$ eV.

\subsection{NanoDCAL (NEGF-DFT)}\label{subsec:NanoDCAL}

NanoDCAL (Nanoacademic Device Calculator) performs self-consistent calculations using the Keldysh nonequilibrium Green’s function(NEGF) formalism in conjunction with a linear combination of atomic orbitals (LCAO), within the framework of density functional theory (DFT)~\cite{NanoDCAL1, NanoDCAL2, Keldysh}. NanoDCAL is used to compute the transmission functions, $\tau(E)$, in consideration of the source-drain voltage $V_\text{ds}$ and the gate voltage $V_{\text{g}}$. Troullier-Martins norm-conserving pseudopotentials were utilized to simulate electron-ionic core interactions. Double-$\zeta$ polarized basis sets were utilized to examine elemental valence electrons. The PBE-GGA exchange-correlation functional was chosen \cite{PBE, GGA}. The equivalent energy cutoff for the grid density was set to $100$~Hartree. The Brillouin zone in reciprocal space was sampled using a $6 \times 1 \times 100$ k-point grid for the electrodes and a $6 \times 1 \times 1$ grid for the central scattering region. The $k$-point grids used for calculating the transmission coefficient and current were $6 \times 1 \times 1$ and $1 \times 1 \times 100$, respectively.

\subsection{Effective gate model}\label{subsec:effective_gate_model}

First-principles calculations in Ref.~[\citenum{FET-AlN}] show that the applied gate voltage $V_g$ induces a linear shift in the chemical potential $\mu$. This shift is more pronounced when $\mu$ lies within the band gap (i.e., $E_V < \mu < E_C$) compared to when it is outside the band gap. Utilizing observations of the band gap shift computed self-consistently from first principles approaches (as seen in Fig.~\ref{Fig:scheme}(d)), the effective gate model, $V^{\mathrm{eff}}_{\mathrm{G}}(V_g)$, is formulated to characterize the alteration in chemical potential caused by the gate voltage. The applied gate voltage shifts the chemical potential from $\mu$ at $V_g=0$ to $\mu+ \mathrm{e}V^{\mathrm{eff}}_{\mathrm{G}}(V_g)$ at $V_g$, as described in Ref.~[\citenum{FET-AlN}]. For the gate architecture shown in Fig.~\ref{Fig:scheme}(c), we use the same parameters as in Ref.~[\citenum{FET-AlN}], with $\alpha_{\mathrm{in}} = 0.83$ and $\alpha_{\mathrm{out}} = 0.33$. When $\mu$ lies within the band gap, the application of $V_g$ shifts the chemical potential by 83\% of e$V_g$, whereas when $\mu$ is outside the band gap, the shift is reduced to 33\% of e$V_g$. 
 
Within the effective gate model, the gate-controlled current is given by
\begin{equation} \label{eq:Landauer}
\begin{aligned}
&I(T, V_g,V_{ds}) =  \\
&\frac{2\mathrm{e}}{h} \int_{-\infty}^{\infty} \left[f(E, T,\mu_L(V_g)) - f(E, T, \mu_R(V_g,V_{ds}))\right] \tau(E) dE,
\end{aligned}
\end{equation}
where $V_{ds}$ is the source-drain bias. The chemical potentials of the left and right leads are $\mu_L(V_g) = \mu + \mathrm{e}  V^{\mathrm{eff}}_{\mathrm{G}}(V_g) $ and $\mu_R(V_g, V_{ds}) = \mu + \mathrm{e} \left [ V^{\mathrm{eff}}_{\mathrm{G}}(V_g) +V_{ds}\right ]$, respectively. The Fermi–Dirac distribution is
\begin{equation}
f \left ( E, T, \mu_{L(R)} \right )  = \frac{1}{e^ {\left [(E-\mu_{L(R))}/(k_B T) \right ]}     + 1}. 
\end{equation}

In the case of minimal bias, Eq.~(\ref{eq:Landauer}) can be approximated as follows 
\begin{equation} \label{eq:Landauer_smallVds}
I(T, V_g,V_{ds}) \approx 
V_{ds}\frac{2\mathrm{e}^{2}}{h} \int_{-\infty}^{\infty} \left[-\frac{\partial f}{\partial E}(E,\mu(V_g)) \right] \tau(E) dE,
\end{equation}
and at $T=0$ K where $-\frac{\partial f}{\partial E}=\delta [E-\mu(V_g)]$, Eq.~(\ref{eq:Landauer}) can be expressed as follows,
\begin{equation} \label{eq:Landauer_T0}
I(T, V_g,V_{ds}) \approx \left\{  G_0  \tau \left[ \mu(V_g) \right]  \right\}V_{ds} .
\end{equation}
where $G_0 =\frac{2\mathrm{e}^2}{h}$ is the unit of quantum conductance. The current in the quantum tunneling regime is approximately $G_0 \tau(\mu)$, where the chemical potential $\mu(V_g)$ is adjustable via $V_g$, produced from electrons near the chemical potential and exhibiting insensitivity to temperature variations.

Similar to the quantum-semiclassical competing parameter $\xi$ defined in Ref.~[\citenum{ZT-TMD}], the total current $I(T,V_g,V_{ds})$ at finite bias can also be decomposed into a quantum-tunneling contribution $I_{\mathrm{QM}}$ and a semiclassical thermionic emission contribution $I_{\mathrm{SC}}(T,V_g,V_{ds})$, where  
\begin{equation} \label{eq:I_QM}
\begin{aligned}
&I_{\text{QM}}(T, V_g,V_{ds}) =  \\
&\frac{2\mathrm{e}}{h} \int_{E_V}^{E_C} \left[f(E, T,\mu_L(V_g)) - f(E, T, \mu_R(V_g,V_{ds}))\right] \tau(E) dE,
\end{aligned}
\end{equation}
and 
\begin{equation} \label{eq:I_SC}
  \begin{aligned}
   &I_{\mathrm{\text{SC}}}(T,V_g,V_{ds})= \\
   &\frac{2\mathrm{e}}{h}\left \{ \int_{-\infty} ^{E_V}+\int_{E_C} ^{\infty} \right \} \left[f(E, T,\mu_L) - f(E, T, \mu_R)\right] \tau(E) dE,
   \end{aligned}
\end{equation}
where $E_V$ and $E_C$ denote the valence-band maximum and conduction-band minimum, respectively. The term $I_{\mathrm{QM}}$ collects contributions from energies within the band gap—i.e., tunneling through a classically forbidden region—whereas $I_{\mathrm{SC}}$ accounts for thermally excited carriers (electrons above $E_C$ or holes below $E_V$) responsible for semiclassical thermionic transport.  

From Eqs.~(\ref{eq:I_QM}) and (\ref{eq:I_SC}), we define a quantum-semiclassical competition parameter \cite{ZT-TMD}, 
\begin{equation} \label{eq:zeta}
   \zeta(T,V_g,V_{ds}) \equiv\frac{I_{\mathrm{SC}}-I_{\mathrm{QM}}}{I_{\mathrm{SC}}+I_{\mathrm{QM}}},
\end{equation}
where $-1 \le \zeta(T,V_g,V_{ds}) \le 1$. A positive $\zeta$ indicates that the total current comprises more semiclassical thermionic current, while a negative $\zeta$ indicates that the total current comprises more quantum tunneling current.

\subsection{\label{sec:asymptotic}Bohr's Correspondence Principle: the asymptotic behavior of Landauer formula}

The Correspondence Principle asserts that as quantum numbers increase, the predictions of quantum physics converge with those of classical physics. In this subsection, we demonstrate that the quantum tunneling current, as represented by the Landauer formula in the quantum transport regime, converges asymptotically to the thermionic current, as represented by Richardson's law in the classical transport regime. In the classical limit, only "hot" electrons with energy above the affinity $\chi$ of the nanojunction potential barrier contribute to the current.

As shown in Ref.~[\citenum{EMT-PW}], the Landauer formula Eq.~(\ref{eq:Landauer}) can be derived from the classical expression $J = n q v$, the quantum field operator, and the nonequilibrium Green's function, respectively, with the transmission coefficient $\tau (E)=\sum_n \tau_n(E)$, where the transmission coefficient contributed from the $n-$th energy band being
\begin{equation*}
\tau_n (E)=\frac{h}{L_x L_y L_z} \sum_n \sum_{k_x,k_y} \sum_{k_z>0} v_{n}(k_z) \delta [E-E(\bf{k})],    
\label{eqn:tauE1}
\end{equation*} 
where the current flows along the positive $z$-axis. If we consider only one band, then the transmission coefficient is given by
\begin{equation}
\tau (E)=\frac{h}{L_z}   \sum _{k_z>0}  v_{z} (k_z) D^{\perp}(E;k_z),
\label{tauE2}
\end{equation}
direction:\begin{equation*}
\begin{aligned}
  D^{\perp}(E;k_z) & =\frac{1}{L_x L_y}\sum_{k_x} \sum_{k_y} \delta [E-E(k_x,k_y,k_z)]  \\
   & =\frac{1}{(2 \pi)^2}\int d k_x \int d k_y \delta [E-E(k_x,k_y,k_z)] \\
   & = \frac{1}{(2 \pi)^2}\int d k_x \int d k_y \delta [E-\{E_z+\frac{k_x^2+k_y^2}{2m}\}],
\end{aligned}
\end{equation*} 
where electrons are free to move in the X-Y plane with $E(\bf{k})=\frac{\hbar^{2}(k_{x}^{2}+k_{y}^{2})}{2 m} +E_z(k_z)$. For each $k_z>0$, subsequent to executing the integration, the partial density of states, $D^{\perp}(E;k_z)$, is
\begin{equation}
D^{\perp}(E;k_z) = \frac{m}{2 \pi \hbar^{2}} \Theta [E-E_{z}(k_z)],
\end{equation}
where $\Theta [E-E_{z}(k_z)]$ is the Heaviside step function. Taking the continuum limit, $\sum_{k_z} \mapsto \frac{L_z}{2 \pi} \int_{-\infty}^{\infty}dk_z$, and we assume that $E_z(k_z) = \frac{\hbar^{2} k_z^{2}}{2m}$ and $v_z(k_z) = \frac{\hbar k_z}{m}$, the transmission function becomes
\begin{equation*}
\begin{aligned}
\tau (E) & =\frac{1}{2 \pi}  \int_{0}^{\infty} k_z \Theta [E-E_{z}(k_z)] dk_z \\
         & = \frac{1}{2 \pi} \int_{0}^{\sqrt{\frac{2 m E}{\hbar ^{2}}}} k_z  dk_z
\end{aligned}
\label{tauE2}
\end{equation*}
In the classical limit, electrons possessing energy $E$ below the electron affinity $\chi$ are restricted from traversing the junction, hence imposing a lower constraint on the aforementioned integral:
\begin{equation}
\begin{aligned}
\tau (E)  ^{\text{asymp.}} & \approx \frac{1}{2 \pi} \int_{\sqrt{\frac{2 m \chi}{\hbar ^{2}}}}^{\sqrt{\frac{2 m E}{\hbar ^{2}}}} k_z  dk_z \\
& = \frac{m}{2 \pi \hbar^{2}} (E-\chi) \Theta (E-\chi) ,
\end{aligned}
\label{eq:tauE3}
\end{equation}
and the Landauer formula corresponding to electrons incident from the left electrode is expressed as follows:
\begin{equation}
\begin{aligned}
J_z & =\frac{2\mathrm{e}}{h} \int  [f(E,T)] \tau(E) ^{\text{asymp.}} dE.
\label{eq:Landaur_assymp1}
\end{aligned}
\end{equation}
%where the upper limit of the integral arises from energy conservation. The lower limit of the integral reflects the fact that, in the classical limit,

For "hot" electrons with energy exceeding the work function $W = \chi - \mu \gg k_B T$, the quantum Fermi–Dirac distribution $f(E,T)$ approximates the classical Boltzmann distribution. Upon substituting the Boltzmann distribution for $f(E,T)$ and Eq.~(\ref{eq:tauE3}) for $\tau(E) ^{\text{asymp.}}$ in the integral above, we obtain
\begin{equation}
J_z =\frac{2\mathrm{e}}{h} \int_{\chi} ^{\infty} \left[\frac{m}{2\pi\hbar^2}(E-\chi)\right ]e^{-(E-\mu)/(k_BT)}  dE    = A T^{2} e^{-\frac{W}{k_B T}}  
\label{eq:Landuaer_assymp2}
\end{equation}
where $A=\frac{\mathrm{e} m k_{B}^{2} }{2 \pi^{2} \hbar^{3}}$ is the Richardson constant. The Landauer formula in the quantum transport regime asymptotically converges to the thermionic emission current characterized by Richardson’s law.  This asymptotic behavior stems from the correspondence principle in quantum mechanics, which guarantees that quantum outcomes converge to classical behavior in the limit of high temperature, large quantum number, and large scale.

In the case of small source-drain bias $V_{ds}$, the net thermionic emission current density, as described by Richardson's law, coming from hot electrons incident from the left and right leads is expressed as
\begin{equation}
J_z = A T^2 e^{-\frac{W}{k_B T}}-A T^2 e^{-\frac{(W+\mathrm{e} V_{ds})}{k_B T}} \approx ( \frac{\mathrm{e} V_{ds}}{k_B})A T e^{-\frac{W}{k_B T}}.
\label{eq:Thermionic_2leads}
\end{equation}

Plotting $J_z/T$ on a $ \log_{10} $ scale as a function of temperature on the $1/T$ scale yields a linear relationship,
\begin{equation}
\log_{10} \left (\frac{J_z}{T} \right )=m \left (\frac{1}{T} \right )+ y_0,
\label{eq:LogJ/T}
\end{equation}
where the slope is given by $m=[\left (\log_{10} \mathrm{e} \right ) / k_B] W$ The slopes of the SC regime (red) correspond to the work function $ W = \chi - \mu(V_g)$, where $W$ denotes the work function, and the y-intercept is given by $y_0=\log_{10} \left[A\left (\mathrm{e} V_{ds}/k_B\right )\right]$. It is important to recognize that the work function, $W(V_g)=\chi-\mu(V_g)$, is dependent on the gate voltage, as $V_g$ influences the chemical potential.

\subsection{Subthreshold swing and Boltzmann tyranny}\label{subsec:current_and_SS}

The subthreshold swing (SS) of a FET quantifies its effectiveness in controlling the current density via the applied gate voltage $V_g$. It is defined as the change in gate voltage required to increase the output current density by one order of magnitude:
\begin{equation}
\text{SS}(T,V_g,V_{ds}) \equiv \ln(10) \left\{ \frac{1}{J(T,V_g,V_{ds})} \frac{dJ(T,V_g,V_{ds})}{dV_g} \right\}^{-1}.
\label{eq:SS}
\end{equation}
Alternatively, within the effective gate model, the subthreshold swing can be expressed as
\begin{equation}
\begin{aligned}
& \text{SS}(T,V_g,V_{ds}) =\left[ \ln(10) \left ( k_{B} T/\mathrm{e}\right ) \right ] \cdot \left [ \frac{dV_{\mathrm{G}}^{\mathrm{eff}}(V_g)}{dV_g} \right ]^{-1} \cdot      \\
&  \frac{ \int_{-\infty}^{\infty} \left [f(E,T,\mu_L(V_g))-f(E,T,\mu_R(V_g,V_{ds})) \right ] \tau(E) dE }{\int_{-\infty}^{\infty} \left \{  \operatorname{sech}^2 \left [  \frac{E-\mu_L(V_g)}{2 k_B T}    \right ]  
- \operatorname{sech}^2 \left [  \frac{E-\mu_R(V_g,V_{ds})}{2 k_B T}    \right ]    \right \}\tau(E) dE },
\end{aligned}
\label{eq:SS_VG}
\end{equation}
where $\left[ \ln(10), (k_B T / \mathrm{e}) \right]$ is the Boltzmann tyranny (BT), the classical limit of the subthreshold swing (SS) in field-effect transistors, while $\left[ \frac{dV_{\mathrm{G}}^{\mathrm{eff}}(V_g)}{dV_g} \right]^{-1}$ representing the gate-controlling efficiency, i.e., $\alpha_{\text{in(out)}}$ in the effective gate model:
\begin{equation}
\begin{split}
\frac{dV_{\mathrm{G}}^{\mathrm{eff}}(V_g)}{dV_g} = 
    \begin{cases}
\alpha_{\mathrm{in}}, & \mathrm{for} ~ \mu(V_g) \in (E_{V},E_{C})  .
\\
    \alpha_{\mathrm{out}}, & \mathrm{otherwise},     
    \end{cases}    
\end{split}
\label{eq:dVGeff/dVg}    
\end{equation} 
where $E_{\mathrm{V}}$ and $E_{\mathrm{C}}$ denote the valence band maximum and conduction band minimum, respectively, which define the edges of the transmission band gap.

When the gate voltage $V_g$ shifts the chemical potential $\mu(V_g)$ into the transmission band gap, a competition arises between quantum tunneling and thermionic emission currents. If the thermionic emission current overwhelmingly dominates the electron transport mechanism, rendering the quantum tunneling current negligible, the subthreshold swing approaches the following limit:
\begin{equation} \label{eq:SSapproximation}
\text{SS} \rightarrow \frac{\ln(10) (k_B T / \mathrm{e})}{\alpha_{\mathrm{in}}},
\end{equation}
which corresponds to the Boltzmann tyranny scaled by the gate control efficiency factor $\alpha_{\mathrm{in}}$.
\section{Results and Discussion}\label{sec:results}

Short-channel effects become increasingly pronounced as transistor dimensions shrink. When the channel length enters the sub-10-nm regime, the dominant transport mechanism can shift from classical thermionic emission to quantum tunneling. Tunneling current grows exponentially with decreasing channel length and is relatively insensitive to temperature, whereas thermionic current is largely length-independent and increases with temperature. However, for FETs scaled to sub-10 nm, the impact of this classical-to-quantum transition on subthreshold swing (SS) and the ON/OFF current ratio remains insufficiently understood.

%Inresponse to Comment 1.4, we add the followings:
This study examines Pt–WSe$_{2}$–Pt nanojunctions subjected to a drain-source bias of $V_{ds} = 50$ mV.  The gate voltage is adjusted within the range of -1.5 V to 1.5 V, while channel lengths of $L_{\mathrm{ch}} = 3$, 6, 9, and 12 nm are analyzed.  The thickness of the gate dielectric space is 8 Å, with a gate dielectrics $\epsilon_r=3.9$ (Fig.~\ref{Fig:scheme}(a) and (c)), while the applied gate voltage is maintained below 1.5 V.  The low voltage level addressed in this manuscript may be insufficient to generate substantial effects on charge screening, interface state filling, and gate leakage.

We compute the transmission coefficient $\tau(E)$ using NEGF-DFT (NanoDCAL) from first-principles approaches. Subsequent to the computation of $\tau(E)$, we evaluate the gate-tunable current and subthreshold swing (SS) via the Landauer formalism combined with the effective gate model described in Subsec.~\ref{subsec:effective_gate_model}. Despite the inherently quantum-mechanical nature of the established framework, it is demonstrated in Subsection~\ref{sec:asymptotic} that the tunneling current, as yielded by the Landauer approach, exhibits asymptotic convergence to the thermionic current. This convergence occurs under specific conditions, namely the long-channel/high-temperature limit, wherein the thermionic current is accurately described by Richardson's law. This observed quantum-classical correspondence furnishes a robust theoretical underpinning for the exploration of the quantum-to-classical crossover phenomena in two-dimensional transition metal dichalcogenide field-effect transistors (2D TMD FETs). The transmission spectra obtained are displayed in Fig.~\ref{Fig:scheme}(b). Figure~\ref{Fig:scheme}(d) illustrates that the transmission band gap exhibits a linear shift in response to the applied gate voltage $V_g$, which serves as the basis for constructing the effective gate model.

As shown in Subsec.~\ref{sec:asymptotic}, quantum-coherent transport described by the Landauer formalism asymptotically approaches classical thermionic emission described by the Richardson–Dushman law. This trend evidences a quantum-to-classical crossover:
(i) Short channels ($L_{\mathrm{ch}}\leq 9$ nm) at $T=300$ K are tunneling-dominated; the conductance is well approximated by $G \approx G_0,\tau_{\mathrm{min}}$, with carriers near the chemical potential within the gap carrying current through the classically forbidden region.
(ii) Long channels ($L_{\mathrm{ch}}=12$ nm) approach semiclassical thermionic emission, where “hot” carriers with energies above the barrier dominate, making the conductance less sensitive to $L_{\mathrm{ch}}$ and more sensitive to temperature.

We analyze transport in Pt–WSe$_2$–Pt monolayer 2D FETs by contrasting the temperature and length dependence of quantum tunneling and thermionic emission. The quantum tunneling current exhibits minimal sensitivity to temperature, yet it decreases exponentially with increasing channel length. In contrast, thermionic current increases with temperature and exhibits minimal dependence on length, adhering to the relationship $J \propto T^2 e^{- \frac{W}{k_{B}T}}$ as outlined in Eq.~(\ref{eq:Landuaer_assymp2}).

\begin{figure} [h]
\includegraphics[width=0.85\linewidth]{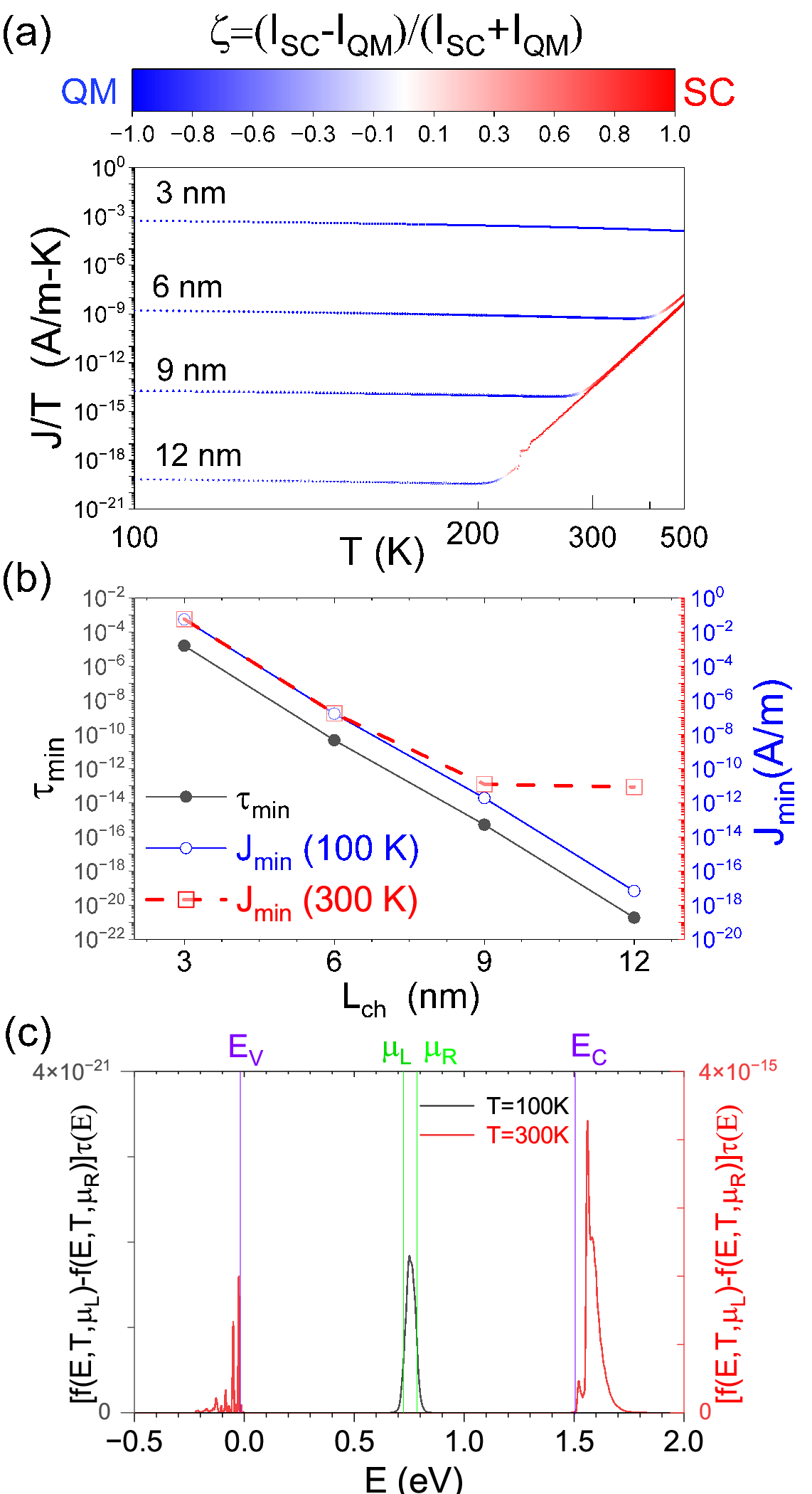} 
\caption{
(a) presents $J/T$ on a $\log_{10}$ scale as a function of temperature ranging from 100 to 500 K plotted against a $1/T$ scale. Note that here $J = J_{\text{min}}$ denotes the current density at $V_g \approx 0.9$ V, which causes the chemical potential $\mu(V_g)$ to align with the center of the transmission band gap. Symbols for each $L_{\mathrm{ch}}$ are color-encoded by $\zeta=(I_{SC}-I_{QM})/({I_{SC}+I_{QM}})$, illustrating the competition between quantum tunneling and thermionic emission currents. The slopes of the SC regime (red) correspond to the work function $ W = \chi - \mu(V_g) $.
(b) displays $\tau_{\text{min}}$ (black solid circles) on the left axis and $J_{\text{min}}$ on the right axis for temperatures of $T=100$ K (blue open circles) and $T=300$ K (red open squares).
(c) displays the integrands, $\left[{(f^{R}-f^{L})\tau(E)}\right ]$, of the Landauer formula as a function of $E$ for $T=100$ K (black line) on the left axis and $T=300$ K (red line) on the right axis.
}
\label{fig:Jmin_tau_Lch}
\end{figure}

In the thermionic transport regime, Eq. (\ref{eq:LogJ/T}) shows a linear relationship between $J/T$ and $1/T$, with a slope of $(\log_{10}e/k_B)W$.

For Pt–WSe$_2$–Pt monolayer nanojunctions, we define $J_{\text{min}}$ as the current density $J(T)$ at $V_g = 0.9$ V, where the gate voltage shifts the chemical potential $\mu(V_g)$ to the midpoint of the transmission band gap $(E_V, E_C)$. At this point, the transmission coefficient reaches its minimum value, denoted $\tau_{\text{min}}$.

Figure \ref{fig:Jmin_tau_Lch}(a) shows the temperature dependence of $J_{\mathrm{min}}(T)/T$ for channel lengths $L_{\mathrm{ch}} =$ 3 , 6, 9, and 12 nm, with the color of each symbol in lines indicating the quantum–semiclassical competition parameter $\zeta = (I_{\text{SC}} - I_{\text{QM}})/(I_{\text{SC}} + I_{\text{QM}})$ (Eq.~ (\ref{eq:zeta})). The results clearly demonstrate a transition in the dominant transport mechanism: at low temperatures, the flat curves and blue color tones correspond to quantum tunneling, whereas at higher temperatures, the increasingly sloped curves and red color tones indicate semiclassical thermionic emission.

It is important to note that $-1 \leq \zeta \leq 1$. Values of $\zeta \leq 0$ signify that quantum tunneling dominates the current, whereas $\zeta \geq 0$ indicates that thermionic emission is the primary transport mechanism. In the semiclassical transport regime, where thermionic emission governs the current, the quantity $J(T)/T$ plotted on a $\log_{10}$ scale against $1/T$ exhibits a linear dependence. The slope of this linear relation is $m = (\log_{10} e)W/k_B$,
where $W(V_g) \approx \chi - \mu(V_g)$ denotes the effective work function at $V_g$, as detailed in Eq.~(\ref{eq:LogJ/T}).

The color-encoded competition parameter $\zeta$ in Fig.~\ref{fig:Jmin_tau_Lch}(a) vividly illustrates the transition between tunneling-dominated (blue tones) and thermionic-emission--dominated (red tones) transport regimes. The turning points with white tones ($\zeta \approx 0$) represent the corssover temperatures at which neither mechanism overwhelmingly dominates.

Moreover, for channel lengths $L_{\text{ch}} = 12$ and 9, the slopes of the $\log_{10}(J/T)$ versus $1/T$ curves are nearly identical, indicating that these FET nanojunctions possess similar barrier heights and effective work functions. In contrast, for $T < 500$~K, the shortest device with $L_{\text{ch}} = 3$~nm remains in the quantum tunneling transport regime.

Figure~\ref{fig:Jmin_tau_Lch}(b) shows $\tau_{\text{min}}$ and $J_{\text{min}}$ at temperatures of 100~K and 300~K on $\log_{10}$ scales for field-effect transistors (FETs) with channel lengths $L_{\mathrm{ch}} = 3$, 6, 9, and 12~nm. The linear dependence of $\tau_{\text{min}}$ on $L_{\mathrm{ch}}$ indicates an exponential decay of the form $\tau_{\text{min}} \propto e^{-\beta L_{\text{ch}}}$. According to Eq.~(\ref{eq:Landauer_T0}), this further implies that $J_{\text{min}} \propto e^{-\beta L_{\text{ch}}}$, which is consistent with transport governed by quantum tunneling.

As a result, at $T = 100$~K, $J_{\text{min}}$ for all channel lengths $L_{\mathrm{ch}} = 3$, 6, 9, and 12~nm remains within the quantum-transport regime. In contrast, for the longest device ($L_{\mathrm{ch}} = 12$~nm) at $T = 300$~K, the transport mechanism transitions from quantum tunneling to semiclassical thermionic emission.

The crossover of the dominant transport mechanism in the $L_{\mathrm{ch}}=12$~nm FET—from quantum tunneling at 100 K to thermionic emission at 300K—is clearly revealed by the integrand $[f^{R}(E,T)-f^{L}(E,T)]\tau(E)$ appearing in Eqs.~(\ref{eq:I_QM}) and (\ref{eq:I_SC}). 
Figure~\ref{fig:Jmin_tau_Lch}(c) shows that at $T=300$~K, transport is primarily governed by semiclassical thermionic emission: the integrand peaks at around $E>E_C$ for electrons and at around $E<E_V$ for holes. 
In contrast, at $T=100$~K, transport is primarily governed by quantum tunneling, as evidenced by a peak in the integrand at the chemical potentials $(\mu_L, \mu_R)$ situated at the center of the band gap.
In this regime, carriers encounter an effective barrier of height approximately $E_C-\mu$ (electrons) or $\mu-E_V$ (holes).

The transition of the dominant transport mechanism in the $L_{\mathrm{ch}}=12$ nm FET from quantum tunneling to thermionic emission is clearly illustrated by the integrand, $\left[{(f^{R}-f^{L})\tau(E)}\right ]$, found in the integral of Eqs.~(\ref{eq:I_QM}) and (\ref{eq:I_SC}). Figure \ref{fig:Jmin_tau_Lch}(c) illustrates that at a temperature of 300 K, transport is primarily governed by semiclassical thermionic emission, with the integrand exhibiting peaks at energies greater than $E_C$ for electrons and less than $E_V$ for holes. At $T = 100$ K, transport is primarily influenced by quantum tunneling, as evidenced by the integrand's peak near the chemical potential within the band gap. This suggests that electrons (or holes) face a barrier with a height approximately equal to $E_C - \mu$ (or $\mu - E_V$).  

\begin{figure*}
\centering
\includegraphics[width=0.85\textwidth]{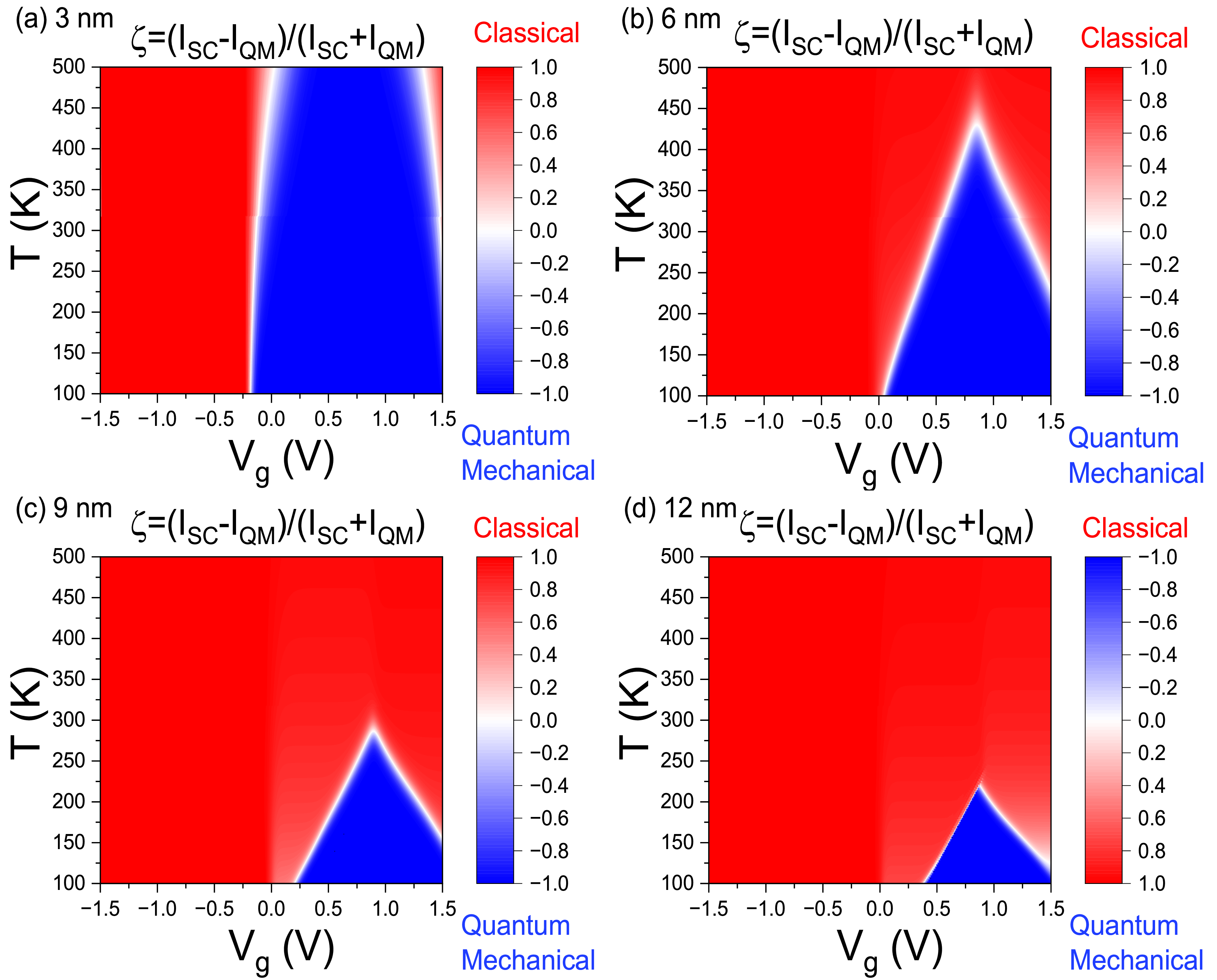} 
\caption{ 
Contour plots of the quantum-classical competition parameter $\zeta$ are presented over a range of  temperature (100–500 K) and gate voltage $V_g$ (–1.5 to 1.5 V) for Pt–WSe$_2$–Pt thermoelectric junctions at $V_{ds}=50$ mV with channel lengths $\mathrm{L}_{\text{ch}}=$ (a) 3 nm, (b) 6 nm, (c) 9 nm, and (d) 12 nm. The rivalry between quantum transport in blue and semi-classical thermionic current in red is represented by $-1 \leq \zeta \leq 1$. 
}
\label{fig:zeta}
\end{figure*}

Figure \ref{fig:zeta} presents the contour plots of the quantum-classical competing parameter $\zeta$ over $T$ and $V_g$ for $L_{ch}=3$, 6, 9, and 12 nm at $V_{ds}=50$ mV.
The white color ($\zeta \approx 0$) delineates the quantum transport regime (blue tone) from the thermionic semi-classical regime, as a direct result of the discussion in Fig.~\ref{fig:Jmin_tau_Lch}(a). At $T=0$ K, no hot electrons or holes are thermally excited outside the band gap to conduct current through semiclassical thermionic emission.  Consequently, we anticipate that the quantum transport regimes (blue tones) will extend across the range of $V_g$, where $\mu(V_g) \in (E_V, E_C)$. The blue-toned regions, excluding the shortest 3 nm FET, exhibit triangle shapes, and it is noteworthy that the tunneling-dominant regimes diminish with rising temperature, leading to a contraction of the green triangular-shaped area as temperature escalates. The tunneling-dominated triangular region ultimately vanishes at temperatures exceeding approximately 250, 325, and 500 K for the 12, 9, and 12 nm FETs, respectively. The exception is the shortest 3 nm FET that remains in the tunneling-dominated region across a broad range of $V_g$ for temperatures below 500 K. Short channels and low temperatures promote the quantum transport regime, whereas long channels and high temperatures prefer the semiclassical electron transport regime. The competitive strength parameter $\zeta$ distinctly marks a quantum-to-semiclassical transition in the $T-V_g$ plane for each $L_{\text{ch}}$.

\begin{figure}
\includegraphics[width=0.8\linewidth]{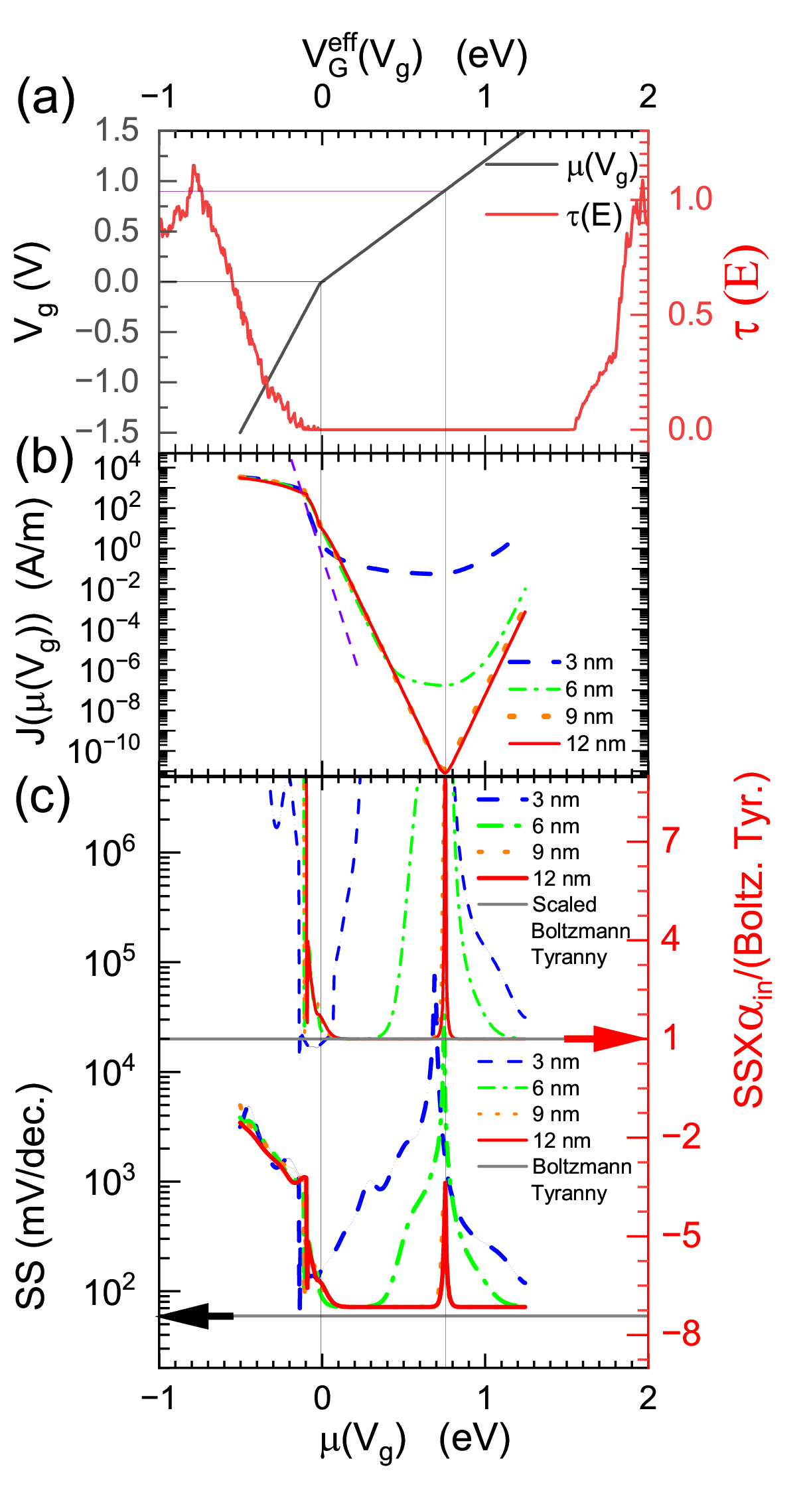}
\caption{ 
(a) shows the transmission coefficient $\tau(E)$ (right vertical axis) as a function of energy with a band gap and illustrates how the gate voltage $V_g$ (left vertical axis), associated with an effective gate voltage $V_{\text{G}}^{\text{eff}}$ (top horizontal axis), shifts the chemical potential $\mu(V_g)$ (bottom horizontal axis) relative to the band gap.
(b) shows the current density $J[\mu(V_g)]$ for Pt–WSe$_2$–Pt junctions with channel lengths $L{\mathrm{ch}} = 3$ nm (blue dashed line), 6 nm (green dash–dotted line), 9 nm (orange dotted line), and 12 nm (red solid line), evaluated at $V_{ds} = 50$ mV and $T = 300$ K.
(c) shows the subthreshold swing, SS$[\mu(V_g)]$, plotted on a $\log_{10}$ scale on the left vertical axis for the same set of junctions, using the color and line-style convention of panel (b). The right vertical axis displays the normalized SS, $\text{SS}\times\big(\alpha_{\text{in}}/\text{BT}\big)$, in linear scale.  The black arrow together with the gray horizontal line indicates the Boltzmann tyranny (BT).
}
\label{fig:J_SS_300K}
\end{figure}

The relationship between the applied gate voltage $V_g$ and the resultant shift of the chemical potential $\mu(V_g)$ in relation to the transmission function $\tau(E)$ is illustrated in Fig.~\ref{fig:J_SS_300K}(a) for the 12 nm Pt–WSe$_2$–Pt nanojunction. As shown by the purple dashed lines, the gate voltage of $V_g = 0.9$ V corresponds to the minimal transmission coefficient, $\tau_{\mathrm{min}}$, which happens at $\mu(V_g) = 0.75$ eV. 

The gate-modulated current density $J(V_g, T, V_{ds})$ for Pt-WSe$_2$-Pt nanojunctions with channel lengths $L_{\mathrm{ch}} = 3$, 6, 9, and 12 nm is plotted on a $\log_{10}$ scale in Fig.~\ref{fig:J_SS_300K}(b) at $T = 300$ K and $V_{ds} = 50$ mV. The vertical purple line indicates the polarity inversion between P-type and N-type operation.  This discussion will concentrate on the P-type regime. With the exception of the 3 nm FET, all devices have a nearly linear dependence of $\log_{10}(J)-V_g$, or $\log_{10}(J)$-$\mu(V_g)$. The linear regime is characterized by a predominant thermionic emission current, resulting in the subthreshold swing (SS) approaching the classical limit, specifically the Boltzmann tyranny (BT) scaled by the gate-control efficiency $\alpha_{\text{in}}$: $\text{SS} \approx \alpha_{in}[\ln(10) k_{B}T/\text{e]}$. Notably, the 3 nm FET exhibits a slope that surpasses the ideal linear region characterized by thermionic emission, suggesting that its subthreshold swing may violate the traditional limitations imposed by the Boltzmann tyranny.

The left axis of Fig.~\ref{fig:J_SS_300K}(c) illustrates the subthreshold swing (SS) in relation to $V_g$. The Boltzmann tyranny at $T=300$ K, $\text{SS}=\ln(10) k_{B}/e=59.23$ $\text{mV~dec}^{-1}$, is indicated on the left axis by the gray line. The nearly flat region yields a subthreshold swing of approximately $SS \approx 72.6$ $\text{mV~dec}^{-1}$, which is accurately represented by the equation $\text{SS}=\alpha_{\text{in}}[\ln(10) k_{B} T/e]$. Normalizing as $\alpha_{\mathrm{in}} \left ( \text{SS}/\text{Boltzmann tyranny} \right )$ demonstrates that SS approaches the classical limit, adjusted by the gate-control factor $\alpha_{\text{in}}$ indicated on the right axis. When $\mu(V_g)$ is a little less than 0, the 3 nm FET generates $\alpha_{\mathrm{in}} \left ( \text{SS}/\text{BT} \right )<1$, demonstrating that strong quantum tunneling has overcome Boltzmann tyranny. The SS overcoming Boltzmann tyranny is attributable to the significant magnitude of the gradient in $\tau(E)$.

\begin{figure*} 
\centering
\includegraphics[width=0.8\textwidth]{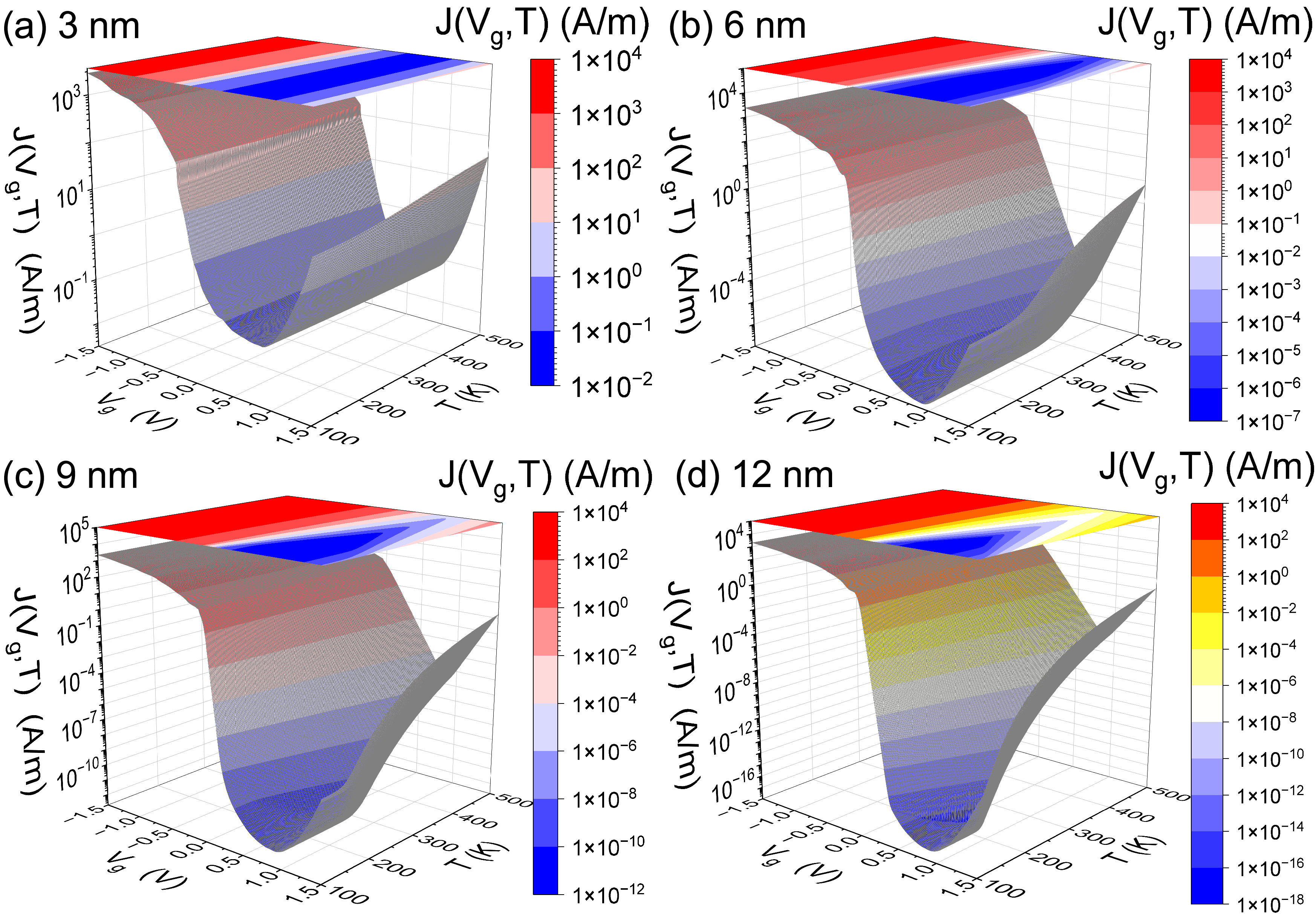}
\caption{ 
Three-dimensional surface plots of the current density $\log_{10}J(V_g, T)$ for Pt–WSe$_2$–Pt nanojunctions as a function of gate voltage $V_g$ (from $–1.5$ to $1.5$ V) and temperature $T$ (from $100$ to $500$ K), evaluated at a fixed drain-source bias of $V_{\mathrm{ds}} = 50$ mV. Results are presented for channel lengths of (a) 3 nm, (b) 6 nm, (c) 9 nm, and (d) 12 nm.
}
\label{fig:J_Vg_T}
\end{figure*}

Figure~\ref{fig:J_Vg_T} illustrates three-dimensional surface plots depicting the current density $J(V_g, T)$ across a gate voltage range of –1.5 to 1.5 V and a temperature range of 100 to 500 K. For $\mu(V_g)$ located in the band gap area, the current density for the 3 nm junction exhibits negligible temperature dependency throughout the range of $100 K \leq T \leq 500 K$, indicating that electron transport is predominantly influenced by quantum tunneling. In the tunneling-dominated regime, the current density decreases exponentially with channel length, $J \propto e^{-\beta L_{\text{ch}}}$, and remains relatively unaffected by temperature. The current density shows a prominent increase, $J \propto e^{-W/k_BT}$, as the temperature increases to approach the thermionic-dominated domain. The above observations align with the established relationships that $J \propto V_{ds} G_0 \tau[\mu(V_g)]$ and $\tau \propto e^{-\beta L_{\text{ch}}}$ in the tunneling-dominated regime and $J \propto T^2 e^{- \frac{W}{k_{\mathrm{B}}T}}$ in the thermionic-dominated regime. 

\begin{figure*} 
\centering
\includegraphics[width=0.8\textwidth]{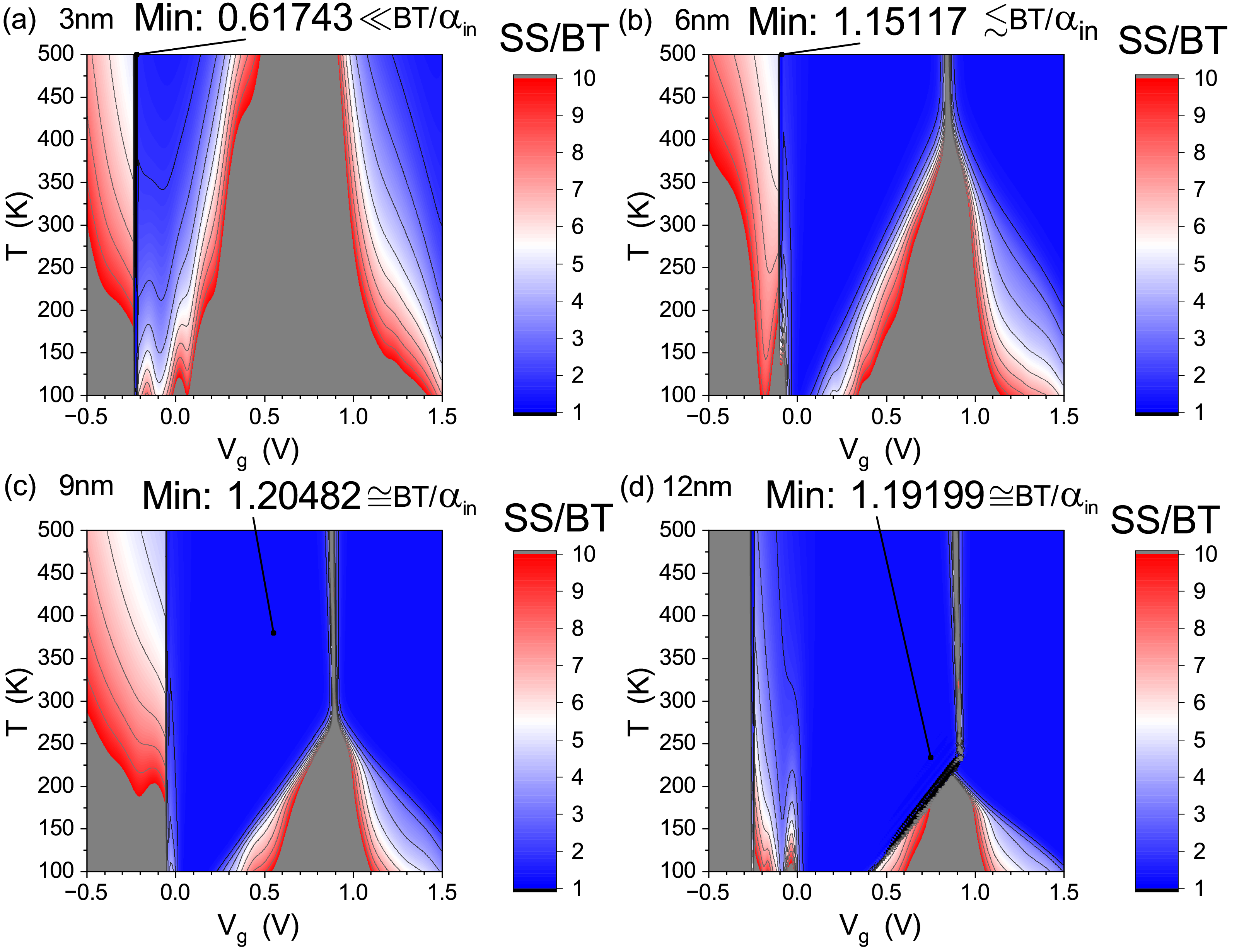} 
\caption{ 
Two-dimensional contour plots of the subthreshold swing (SS), normalized to the values of Boltzmann tyranny (BT), for Pt–WSe$_2$–Pt nanojunctions as a function of gate voltage $V_g$ (ranging from –0.25 to 1.5 V) and temperature $T$ (ranging from 100 to 500 K), evaluated at a fixed drain–source bias of $V_{\mathrm{ds}} = 50$ mV. The results are shown for channel lengths of (a) 3 nm, (b) 6 nm, (c) 9 nm, and (d) 12 nm.
The value and location of the lowest $\text{SS}/\text{BT}$ value in the P-type zone is shown as Min at the top of every T$-V_g$ contour map and compared with the classical limit $\text{BT}/\alpha_{in}$, where $\alpha_{\text{in}}=$ 0.83.
%Note that $\text{BT}=\ln(10)(k_B T/\text{e})$ varies with temperature and is approximately 60 $\text{mV~dec}^{-1}$ at T=300 K.
}
\label{fig:SS_Vg_T}
\end{figure*}

Figure~\ref{fig:SS_Vg_T} presents two-dimensional contour plots of the subthreshold swing (SS) normalized to the Boltzmann tyranny (BT), $\text{SS}(V_g, T)/\text{BT}(T)$, over a gate voltage range of –0.25 to 1.5 V and a temperature range of 100 to 500 K for 2D FETs with channel lengths $L_{\text{ch}}=3$, 6, 9, and 12 nm. At the top of each subgraph, the minimum value of $\text{SS}/\text{BT}$ in the P-type zone is displayed.
Note that $\text{BT(T)}=\ln(10)(k_B T/\text{e})$ varies with temperature and is approximately 60 $\text{mV~dec}^{-1}$ at T=300 K.

The minimum $\text{SS}/\text{BT}$ value in the P-type zone is displayed at the apex of each graph and is compared with $\text{BT}/\alpha_{\text{in}}$, where $\alpha_{\text{in}} = 0.83$ representing the gate control efficiency. For $L_{\text{ch}} \geq 9$, the minimum $\text{SS}/\text{BT}$ values converge to $\text{BT}/\alpha_{\text{in}}$, signifying that the current ideally approaches the thermionic limit. The scaling behavior of the gate control efficiency $\alpha_{\text{in}}$ demonstrates that our theoretical framework is universally applicable when the transport mechanism functions within the semi-classical thermionic transport regime. The subthreshold swing can be improved by reducing the thickness of the dielectirc spacers between the channel material and the gate electrode. It is important to note that the two-dimensional geometry of the channel material in 2D FETs enables researchers to fabricate extremely thin dielectric spacers, thereby optimizing gate control efficiency. Consequently, some experimental groups report subthreshold swing (SS) values approaching 60 $\text{mV~dec}^{-1}$. 

As the channel length decreases to 3 nm, the minimum value of the subthreshold swing approaches around $0.617 \alpha_{\text{in}}\times \text{BT}$. The black area in Fig.~\ref{fig:SS_Vg_T}(a) signifies that quantum transport may not conform to the classical limit dictated by Boltzmann tyranny, even with an imperfect gate control efficiency of $\alpha_{\text{in}}=0.83$. The blue regions indicate low values of subthreshold swing and correlate with a linear $\log_{10}(J)$–$V_g$ relationship, signifying effective gate modulation efficiency.

The expansion of these blue regions with increasing temperature indicates enhanced gate control in thermionic-dominated transport regimes. A similar expansion is observed with increasing channel length, but this effect saturates beyond 9 nm, consistent with the saturation of thermionic enhancement discussed earlier.

\begin{figure*} 
\centering
\includegraphics[width=0.7\textwidth]{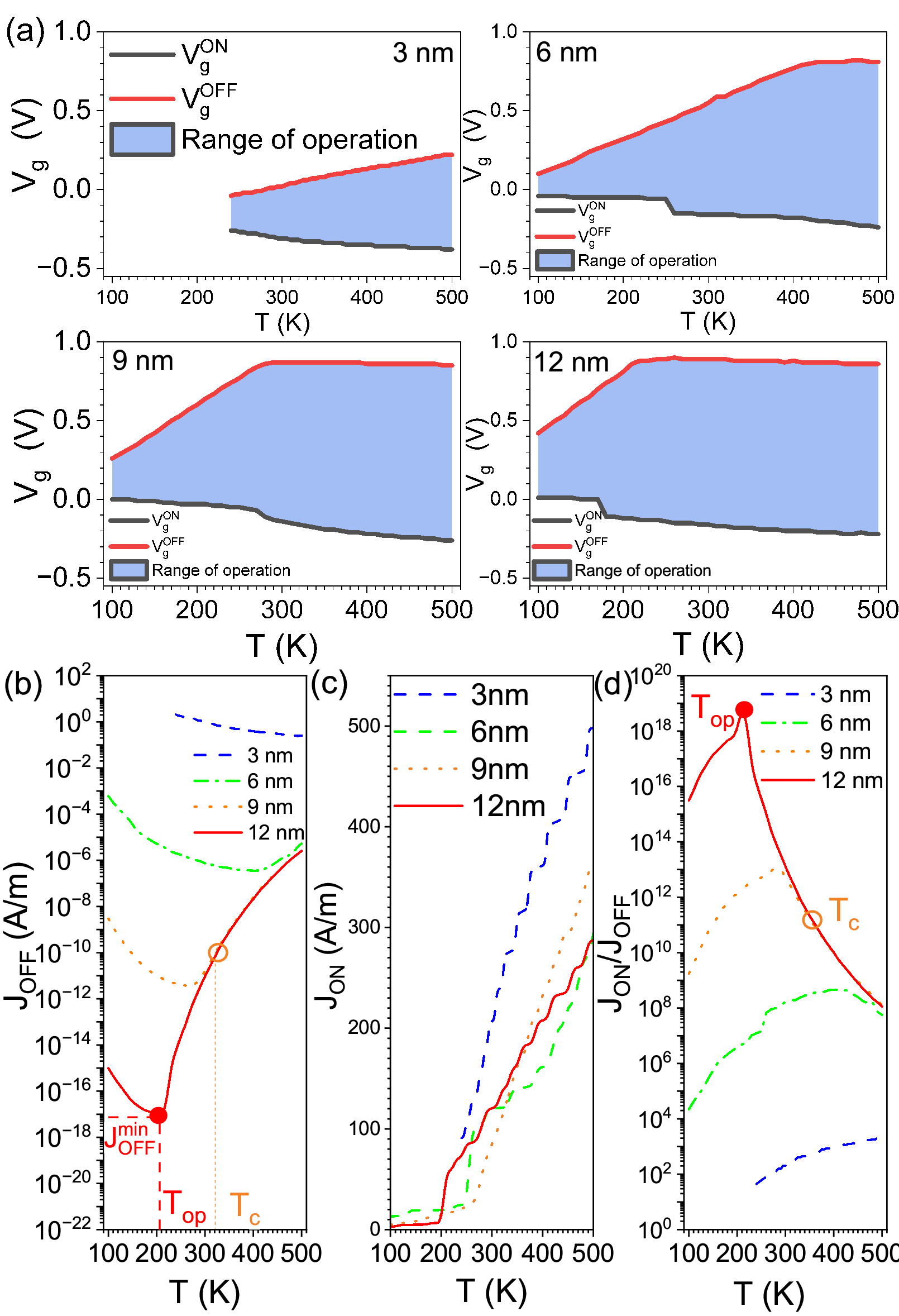}
\caption{ 
(a) The optimal operating range of Pt–WSe$_2$–Pt nanojunctions as field effect transistors with channel lengths $L{\text{ch}} = 3$, 6, 9, and 12 nm at $V_{ds} = 50$ V, defined by the condition $\text{S.S.} \leq 3 \times (\text{Boltzmann tyranny})$.
(b) shows the OFF-state current density, $J_{\mathrm{OFF}}$, for $L_{\mathrm{ch}} = 3$ nm (blue dashed), 6 nm (green dash–dotted), 9 nm (orange dotted), and 12 nm (red solid). The red solid circle denotes the optimized temperature, $T_{\mathrm{op}}$, at which the OFF current reaches its minimum, $J_{\mathrm{OFF}}^{\mathrm{min}}$, indicating the optimal turn-off condition for the 12 nm junction. All $J_{\mathrm{OFF}}$ curves eventually converge to the same trajectory. The open orange circle indicates the critical temperature, $T_{\mathrm{c}}$, at which the 9 nm curve merges with the 12 nm curve.
(c) shows the ON-state current density, $J_{\mathrm{ON}}$. (d) shows the ON/OFF current ratio, $J_{\mathrm{ON}}/J_{\mathrm{OFF}}$. Both panels use the color and line-style conventions of panel (b).
}
\label{fig:RangeOP_T}
\end{figure*}

For each temperature $T$, the criterion $\text{S.S.} = 3 \times [\ln(10), k_{\mathrm{B}}T/e]$ delineates the nearly linear section of the $\log_{10} J$–$V_g$. In this linear regime, the $\log_{10} J$–$V_g$ graph facilitates the precise identification of the OFF- and ON-state current densities, $J_{\mathrm{OFF}}$ and $J_{\mathrm{ON}}$, along with their respective gate voltages, $V_g^{\mathrm{OFF}}$ and $V_g^{\mathrm{ON}}$, collectively delineating the optimal operating window at that temperature. The optimal operating regime is presented in the shaded area of Fig.~\ref{fig:RangeOP_T}(a). The optimal operational range expands as the channel length increases from 3 nm to 9 nm; however, it levels off beyond 9 nm, suggesting diminishing returns in enhancing gate-control efficiency for longer channels. 

Figure~\ref{fig:RangeOP_T}(b) shows the OFF current $J_{\mathrm{OFF}}$ for FETs with $L{\mathrm{ch}} = 3$, 6, 9, and 12 nm. As temperature rises, $J_{\mathrm{OFF}}$ initially declines, attains a minimal value $J_{\text{OFF}}^{\text{min}}$, and subsequently increases—establishing an optimal temperature $T_{\text{op}}$ (shown by a red solid circle for 12 nm) at which turn-off is maximally efficient. The values of $J_{\text{OFF}}^{\text{min}}$ and $T_{\text{op}}$ rise as the channel length decreases. For $T < T_{\text{op}}$, the FETs mostly function in the tunneling-dominated regime, with the OFF current density $J_{\text{OFF}} \approx V_{ds}G_o \tau$, where $\tau \propto e^{-\beta L_{\text{ch}}}$. For $T > T_{\text{op}}$, the OFF-state current $J_{\text{OFF}}$ increases with rising temperature. The increase in $J_{\text{OFF}}$ arises from the growing contribution of thermally excited hot electrons to the thermionic emission current. We observe that $J_{\text{OFF}}$ curves of all FETs tend to converge onto the same trajectory. Consequently, a critical temperature, $T_{\text{c}}$, can be defined as the point at which a given $J_{\text{OFF}}$ curve merges with this trajectory. For $T > T_{\text{c}}$, $J_{\text{OFF}}$ is dominated by thermionic emission, and the FET operates in the semiclassical transport regime.

At room temperature, long-channel FETs ($L_{\text{ch}} \ge 9$ nm) have an optimal temperature $T_{\text{op}}$ below 300 K, indicating that the OFF current $J_{\text{OFF}}$ is thermionic-dominated and largely independent of $L_{\text{ch}}$. By contrast, short-channel FETs ($L_{\text{ch}} < 9$ nm) exhibit $T_{\text{op}} > 300$ K, signifying that $J_{\text{OFF}}$ is governed by quantum tunneling and increases as the channel shortens, according to $J \propto e^{-\beta L_{\text{ch}}}$. In this regime, quantum tunneling generates substantial “leakage” current, elevating $J_{\text{OFF}}$ and complicating reliable turn-off in ultra-short devices.

Figure~\ref{fig:RangeOP_T}(c) shows the ON current $J_{\mathrm{ON}}$ for FETs with $L{\mathrm{ch}} = 3$, 6, 9, and 12 nm. As $|V_g^{\text{ON}}|$ increases with temperature, the corresponding $|\tau(E)|$ also increases, leading to a rise in $J_{\mathrm{ON}}$ with increasing $T$. The 3 nm FET exhibits a larger $J_{\text{ON}}$ because its $V_g^{\text{ON}}$ is particularly small, resulting in a relatively larger $\tau(E)$ compared to longer channel lengths. Figure~\ref{fig:RangeOP_T}(d) shows the ON–OFF current ratio, $J_{\mathrm{ON}}/J_{\mathrm{OFF}}$, for $L_{\text{ch}}=3$, 6, 9, and 12 nm. Because $J_{\text{OFF}}$ spans orders of magnitude more than $J_{\text{ON}}$, the ratio $J_{\mathrm{ON}}/J_{\mathrm{OFF}}$ essentially exhibits the opposite trend to $J_{\text{OFF}}$.

%Inresponse to Comment 2.1, we add the followings:
Substantial experimental efforts have been dedicated to realizing near-ideal subthreshold swing (SS) values approaching the classical thermionic-emission limit of 60 mV~dec$^{-1}$ ~\cite{Batool_Small_2023,Illarionov_Nature_Comm_2020,Mitta_2D_materials_2021}. Several research groups have successfully enhanced SS performance by exploiting the unique advantages of two-dimensional device architectures. For example, monolayer MoS$_2$/h-BN FETs have demonstrated SS values of approximately 75~mV~dec$^{-1}$~\cite{Fang_AdvFuncMater_2019}, while further improvement to around 69~mV~dec$^{-1}$ has been achieved by vertically confining the MoS$_2$ channel between a top h-BN layer and bottom h-BN gate dielectrics~\cite{Quoc_2DMaterials_2018}. Near-ideal SS values have also been reported in self-assembled Pt/MoS$_2$ Schottky junctions~\cite{Jingli_Wang_ScienceAdv_2021}, WSe$_2$ transistors gated by ionic liquids~\cite{Prakash_ACSNano_2017}, and p-type WSe$_2$ FETs with optimized channel thickness~\cite{Ali_AdvElectMater_2024}.

It is important to emphasize, however, that these experimental demonstrations operate predominantly in the diffusive transport regime, where the channel length exceeds the electron–phonon mean free path. In contrast, our study focuses on the quasi-ballistic transport regime, in which the channel length is shorter than the mean free path and scattering is strongly suppressed. Consequently, the theoretical results presented here provide a forward-looking guideline for future device engineering aimed at scaling channel lengths to the sub-10-nm regime and minimizing scattering-induced resistance.

\section{Conclusion}\label{sec:conclusion}

%As channel lengths shrink into the sub-10-nm regime, short-channel effects emerge and the interplay between classical and quantum transport begins to set scaling limits in 2D FETs. Yet how this interplay governs device metrics below 10 nm remains insufficiently established. In this study, we have developed a unified, first-principles framework that identifies the crossover between quantum-coherent tunneling and semiclassical thermionic emission in monolayer Pt–WSe$_2$–Pt FETs with channel lengths of 3, 6, 9, and 12 nm. Transmission spectra $\tau(E)$ are calculated using NEGF–DFT (NanoDCAL). The effective gate model, in conjunction with the Landauer formalism, is utilized to assess gate-dependent current density $J(V_g,T)$, subthreshold swing (SS), OFF current, ON/OFF ratio, and operating windows across the range of $V_g \in [-1.5,1.5]$ V and $T \in [100,500]$ K at a drain-source voltage of $V_{\mathrm{ds}} = 50$ mV.

We show that, in the long-channel/high-temperature limit, the Landauer formalism asymptotically reduces to Richardson’s thermionic current. We further introduce a competition parameter, $\zeta \equiv (I_{\mathrm{SC}}-I_{\mathrm{QM}})/(I_{\mathrm{SC}}+I_{\mathrm{QM}})$, to delineate the transition between semiclassical thermionic emission and quantum tunneling. This provides a practical metric for characterizing the semiclassical-to-quantum crossover as channel length decreases and temperature is lowered in 2D FETs.

In the quantum-tunneling--dominated regime, the current is primarily contributed by electrons whose energies lie near the chemical potentials, as characterized by $\tau(\mu_{L(R)})$. Consequently, the current scales approximately as $G_0 \tau(\mu) V_{ds}$. In contrast, in the classically dominated regime, the current is largely determined by electrons with energies slightly above the barrier height.
As described in Eq.~(14), $\log_{10}(J/T)$ versus $1/T$ is linear with a slope $m=[(\log_{10} e) /k_B]W$ related to the work function. The transition temperature separating quantum and classical transport decreases with increasing channel length, as shown in Fig.~\ref{fig:Jmin_tau_Lch}(a). This classical-to-quantum crossover arises from the interplay between the semiconductor transmission profile $\tau(E)$ and the energy-dependent electron injection statistics $(f^R - f^L)\tau(E)$, as illustrated in Fig.~\ref{fig:Jmin_tau_Lch}(c). Consequently, variations in TMD channel configurations yield qualitatively similar classical-to-quantum crossover behavior, apart from potential shifts and scaling in the $J_{ds}$--$V_g$ characteristics. Different combinations of metallic electrodes and two-dimensional channel materials modify both the barrier height and width, as well as the chemical potential alignment, through interface charge transfer and Fermi-level pinning. Variations in EOT and $\epsilon_{r}$ modify the gate-control efficiency $\alpha_{\text{in}}$ within the effective gate model, resulting in a universal scaling behavior of the subthreshold swing, $\text{SS}/\alpha_{\text{in}}$, in the thermionic transport regime. As illustrated in Fig.~\ref{fig:J_SS_300K}(c), this normalized subthreshold swing approaches the fundamental limit 
$\text{SS}/\alpha_{\text{in}} \approx \ln(10)\ k_B T/e$.

Two characteristic temperatures, $T_{\text{op}}$ and $T_{\text{c}}$ can be identified. The optimization temperature $T_{\text{op}}$ is defined as the temperature at which $J_{\text{OFF}}$ reaches its minimum value, effectively achieving optimal transistor turn-off. For $L_{\text{ch}} < 9$ nm FETs, $T_{\text{op}}$ is smaller than the room temperature and  $J_{\text{OFF}}$ is dominated by quantum tunneling. Short-channel quantum tunneling leads to leakage current, thereby increasing $J_{\text{OFF}}$ and complicating the process of turning off the transistor current in ultra-short devices. For $L_{\text{ch}} \gtrsim 9$ nm FETs,  $J_{\text{OFF}}$ is mostly determined by the thermionic emission current. In this regime, the subthreshold swing approaches ($\text{BT}/\alpha_{\text{in}}$). where BT and $\alpha_{\text{in}}$ represents the Boltzmann tyranny and gate-controlling efficiency, respectively.

With the continued rise in temperature, a second characteristic temperature, $T_{\text{c}}$, appears, indicating the point at which the current is predominantly influenced by thermionic emission. $T_{\text{c}}$ denotes the threshold beyond which thermionic emission predominantly prevails, and the SS nears $\frac{\text{BT}}{\alpha_{\text{in}}}$. The optimal operational range expands as $L_{\text{ch}}$ increases from 3 to 9 nm, but exhibits saturation at 12 nm FET. This suggests that a channel length of $L_{\text{ch}} \approx 10$ nm is nearly optimal for the performance of 2D FETs. Our findings establish criteria for defining the ideal operating temperature and gate-voltage window in the downscaling of 2D FETs, and they locate the crossover point at which the quantum tunneling current becomes comparable to the semiclassical thermionic emission current.

\section*{Data Availability Statement}

The data that support the findings of this study are available from the corresponding author upon reasonable request.

\begin{acknowledgments}
The authors gratefully acknowledge financial support from the National Science and Technology Council (NSTC), Taiwan, under Grants No. NSTC-111-2112-M-A49-032- and No. NSTC-114-2112-M-A49-034-. This work was also supported by the Advanced Semiconductor Technology Research Center through the Featured Areas Research Center Program, within the framework of the Higher Education Sprout Project funded by the Ministry of Education (MOE), Taiwan. Additional support was provided by the NSTC T-Star Center Project Future Semiconductor Technology Research Center, under Grants No. NSTC-114-2634-F-A49-001- and No. NSTC-113-2112-M-A49-037-. The authors also thank the National Center for High-Performance Computing (NCHC) for providing computational and storage resources.
\end{acknowledgments}

%\appendix

%\section{Appendixes}

%\section{A little more on appendixes}

\bibliography{ref}% Produces the bibliography via BibTeX.

\end{document}